\documentclass[aps,preprint,prd,nofootinbib]{revtex4-1}
\pdfoutput=1	
\usepackage[utf8]{inputenc}
\usepackage{booktabs}
\usepackage{slashed}
\usepackage{indentfirst}
\usepackage[pdftex]{graphicx}
\usepackage{epstopdf}
\usepackage{amsfonts,amssymb,amsmath}
\usepackage{psfrag}

\usepackage{multirow}
\usepackage{hyperref}
\usepackage{mathrsfs} 
\usepackage{url}
\usepackage{appendix}
\usepackage{subfigure}
\usepackage{color}
\usepackage{soul}
\usepackage{ulem}

\newcommand{\ud}{{\rm{d}}}
\newcommand{\ui}{{\rm{i}}}
\newcommand{\ue}{{\rm{e}}}
\newcommand{\GeV}{{\ \rm{GeV}}}

\begin{document}

\title{Strong Decays of the Orbitally Excited Scalar $D^{*}_{0}$ Mesons}

\author{Xiao-Ze Tan$^{[1]}$, Tianhong Wang$^{[1]}$\footnote{thwang@hit.edu.cn}, Yue Jiang$^{[1]}$, Si-Chen Li$^{[1]}$, \\
	Qiang Li$^{[1,3]}$, Guo-Li Wang$^{[1]}$, Chao-Hsi Chang$^{[2,3]}$}

\address{$^1$Department of Physics, Harbin Institute of
	Technology, Harbin, 150001, People's Republic of China \\
	$^2$CCAST(World Laboratory), P.O. Box 8730, Beijing 100080, People's Republic of China \\
	$^3$Institute of Theoretical Physics, Chinese Academy of Sciences, \\
	P.O. Box 2735, Beijing 100080, People's Republic of China}

\baselineskip=20pt

\vspace*{0.5cm}

\begin{abstract}

We calculate the two-body strong decays of the orbitally excited scalar mesons $D_0^*(2400)$ and $D_J^*(3000)$ by using the relativistic Bethe-Salpeter (BS) method. $D_J^*(3000)$ was observed recently by the LHCb Collaboration, the quantum number of which has not been determined yet. In this paper, we assume that it is the $0^+(2P)$ state and obtain the transition amplitude by using the PCAC relation, low-energy theorem and effective Lagrangian method. For the $1P$ state, the total widths of $D_0^*(2400)^{0}$ and $ D_0^*(2400)^+$ are 226 MeV and 246 MeV, respectively. With the assumption of $0^+(2P)$ state, the widths of $D_J^*(3000)^0$ and $D_J^*(3000)^+$ are both about 131 MeV, which is close to the present experimental data. Therefore, $D_J^*(3000)$ is a strong candidate for the $2^3P_0$ state.

 \vspace*{0.5cm}


\end{abstract}

\maketitle


\section{INTRODUCTION}

In recent years, many new charmed mesons have been discovered experimentally, including lots of orbitally high excited states. For example, in 2004, the FOCUS Collaboration \cite{Focus2004} and the Belle Collaboration \cite{Belle2004} observed the $D_0^{*}$, which is the $1P$ scalar and has been studied widely and carefully
\cite{Close2005,Godfrey2005,zhaoqiang2008}. In 2013, the LHCb collaboration announced several new charmed structures, including the $D_J(3000)$ and $D_J^*(3000)$ \cite{LHCb2013}. The $D_J(3000)$ was observed in the $D^{*}\pi$ mass spectrum. Its mass and width are $2971.8\pm 8.7 \ \mathrm{MeV}$ and $188.1 \pm 44.8\  \mathrm{MeV}$, respectively. Spin analysis indicates that $D_J(3000)$ has an unnatural parity, and the assignments of $2P(1^+)$, $3S(0^-)$ and $1F(3^+)$ etc. have been discussed\cite{Sun2013,Lu2014,YuGL2015,Godfrey2016}. Our previous study favored the broad $2P(1^+)$ assignments\cite{lsc2018}.

The $D_J^*(3000)$ is observed in the $D\pi$ mass spectrum, whose mass and width are
\begin{eqnarray}
\begin{aligned}
M&_{D_J^*(3000)}=3008.1\pm 4.0 \ \mathrm{MeV}, \\
\Gamma &_{D_J^*(3000)}=110.5 \pm 11.5\  \mathrm{MeV}.
\end{aligned}
\end{eqnarray}
The parity of this particle is still uncertain in present experiments. From its decay mode of $D\pi$, many authors treat it as a natural parity particle. Considering that its mass is around $3000$ MeV, the assignments of $2^3P_0$, $1^3F_4$, $3^3S_1$, $1^3F_2$ and $2^3P_2$ are possible \cite{Zhu2017}. Different models give the theoretical predictions of their masses and we summarized them in Table \ref{massspectrum}.  The OZI-allowed strong decays with these possible assignments also have been studied by several models, and the results are summarized in Table \ref{decayduibi}.

\begin{table}[htb!]
	\renewcommand\arraystretch{0.95}
	\caption[massspectrum]{Several natural parity candidates of $D_J^*(3000)^0$ (MeV) }\label{massspectrum}
	\vspace{0.3em}\centering
	\begin{tabular}{ccccccc}
		\toprule[1.5pt]
		{$J^P$} & {$n^{2S+1}J_L$}	& Godfrey1985\cite{Godfrey1985}   & Pierro2001\cite{Pierro2001} & Ebert2009\cite{Ebert2009} & Sun2013\cite{Sun2013} & Godfrey2016\cite{Godfrey2016} \\
		\midrule[1pt]
		\multirow{2}{*}{$0^+$} & $1 ^3P_0$ & 2400 & 2377 & 2466 & 2398 & 2399 \\
		{ } 					& $2 ^3P_0$ & -    & 2949 & 2919 & 2932 & 2931 \\
		{$1^-$}					& $3 ^3S_1$ & -    & 3226 & 3096 & 3111 & 3110 \\
		{$2^+$}					& $2 ^3P_2$ & -	   & 3035 & 3012 & 2957 & 2957 \\
		\multirow{2}{*}{$3^-$} & $1 ^3D_3$ & 2830 & 2799 & 2863 & 2833 & 2833 \\
		{ }						& $2 ^3D_3$ & -    & -    & 3335 & 3226 & 3226 \\
		{$4^+$}					& $1 ^3F_4$ & 3110 & 3091 & 3187 & 3113 & 3113 \\
		\bottomrule[1.5pt]
	\end{tabular}
\end{table}

\begin{table}[htb!]
	\renewcommand\arraystretch{0.95}
	\caption[decayduibi]{Decay widths of $D_J^*(3000)^0$ with different assignments (MeV) }\label{decayduibi}
	\vspace{0.3em}\centering
	\begin{tabular}{cccccccc}
		\toprule[1.5pt]
		{\quad $n^{2S+1}L_J$ \quad}	& Mode&\quad Sun\cite{Sun2013} \quad & \quad Yu\cite{YuGL2015} \quad & \quad L{\"{u}}\cite{Lu2014} \quad  & \quad Song\cite{Song2015} \quad &   Godfrey\cite{Godfrey2016} \ \\
		\midrule[1pt]
		\multirow{4}{*}{$3 ^3S_1$} & $D\pi$ & 0.91 & 5.45 & 14.0 & 13.5 & 3.21 \\
		{ } 					  & $D^*\pi$ & 3.5  & 4.85 & 19.4 & 25.7 & 5.6  \\
		
		{ }	  				   & Total    & 18.0  & 87.2 & 158.0 & 103.0 & 80.4 \\
      \midrule[0.7pt]
		\multirow{4}{*}{$2 ^3P_0$} & $D\pi$ & 49   & 35.9 & 83.5 & 72.5 & 25.4 \\
		{ } 					  & $D^*\pi$ & -  & -    & -    & -    & -    \\
		
		{ }	  				   & Total    & 194  & 224.5 & 639.3 & 298.4 & 190  \\
     \midrule[0.7pt]
		\multirow{4}{*}{$2 ^3P_2$} & $D\pi$ & 1.8  & 5.0 & 1.92 & 1.46 & 5.0 \\
		{ } 					 & $D^*\pi$ & $8.1\times 10^{-3}$  &  17.8  & 11.89 & 0.12 & 17.1  \\
		
		{ }	  				   & Total    & 47.0  & 174.5 & 110.5 & 68.9 & 114  \\
      \midrule[0.7pt]
		\multirow{4}{*}{$1 ^3F_2$} & $D\pi$ & 16  & 18.8 & 28.6 & 26.1 & 23.1 \\
		{ } 					  & $D^*\pi$ & 13   & 15.7 & 21.0 & 18.8 & 18.5  \\
		
		{ }	  				   & Total    & 136  & 116.4 & 342.9 & 222.0 & 243 \\
       \midrule[0.7pt]
		\multirow{4}{*}{$1 ^3F_4$} & $D\pi$ & 1.2  & 21.3 & 9.96 & 4.97 & 15.8 \\
		{ } 					  & $D^*\pi$ & 1.8  & 14.1 & 9.41 & 5.31 & 15.2  \\
		
		{ }	  				   & Total    & 39 & 102.3 & 103.9 & 94.5 & 129 \\
		\bottomrule[1.5pt]
	\end{tabular}
\end{table}

Since the parity is conserved in strong decays, the $D^{*}\pi$ channel is forbidden for the $^3P_0$ states. In Table \ref{decayduibi}, all assignments except $2^3P_0$ have both $D\pi$ and $D^{*}\pi$ decay modes and most calculations give the similar decay widths of these two channels. However, $D_J(3000)$ was only found in $D^{*}\pi$ spectrum, while $D_J^*(3000)$ only in $D\pi$ spectrum \cite{LHCb2013} in LHCb experiment. The theoretical results that $D_J^*(3000)$ has similar decay widths of $D\pi$ and $D^{*}\pi$ modes are not consistent with present experimental data. Thus, the assignment of $2^3P_0$ for $D_J^*(3000)$ is more reasonable and some recent researches also favor this assignment \cite{Gupta2018}.

We also note that the theoretical predictions for the total widths of $D_J^*(3000)$ as the $2^3P_0$ state are larger than the experimental data. It can be explained that the estimated decay width by calculating the OZI-allowed strong decays is sensitive to its mass and there are divergences of the mass values between the preliminary detection of the $D_J^*(3000)$ with the present theoretical predictions. In our previous work, we have found that the excited states have large relativistic corrections, so non-relativistic or semi-relativistic models may give large uncertainties. This conclusion can be obtained from the results in Table \ref{decayduibi}: all the assignments of $D_J^*(3000)$ are highly excited states and The corresponding results vary from different methods. For example, the total width for the $3^3S_1$ case ranges from $18$ to $158$ MeV, which shows large divergences between different methods.

Thus, we treat $D_J^*(3000)$ as the second excited state of P-wave scalar meson ($2^3P_0$), and calculate its OZI-allowed two-body strong decays, trying to find out if it is consistent with the LHCb results.
We use the improved Bethe-Salpeter (BS) method \cite{BS1951,Salpeter1952} which contains the relativistic corrections \cite{WangGL2004,Wangth2013,LiQ2017}. In all possible channels, there is a light meson in the final state. We use the reduction formula, Partially Conserved Axial-vector Current(PCAC) relation, and low-energy theorem to deal with the case when the light final meson is a pseudo-scalar. This approach cannot be applied to the channels containing a light vector meson. So, we also adopt the effective Lagrangian method \cite{Wangth2017}.

The rest content of this paper is organized as follows. In Sec.~\uppercase\expandafter{\romannumeral2}, we derive the form of transition amplitudes with BS method and show the details of the effective Lagrangian method. In Sec.~\uppercase\expandafter{\romannumeral3}, we give the numerical results of OZI-alowed two-body strong decays of $D_0^*(2400)$ and $D_J^*(3000)$, and compare them with other researches. Summary and conclusion are presented in Sec.~\uppercase\expandafter{\romannumeral4}.
\section{Two-body Strong Decay}

\begin{figure}[htb]
	\centering
	\includegraphics[scale=0.41]{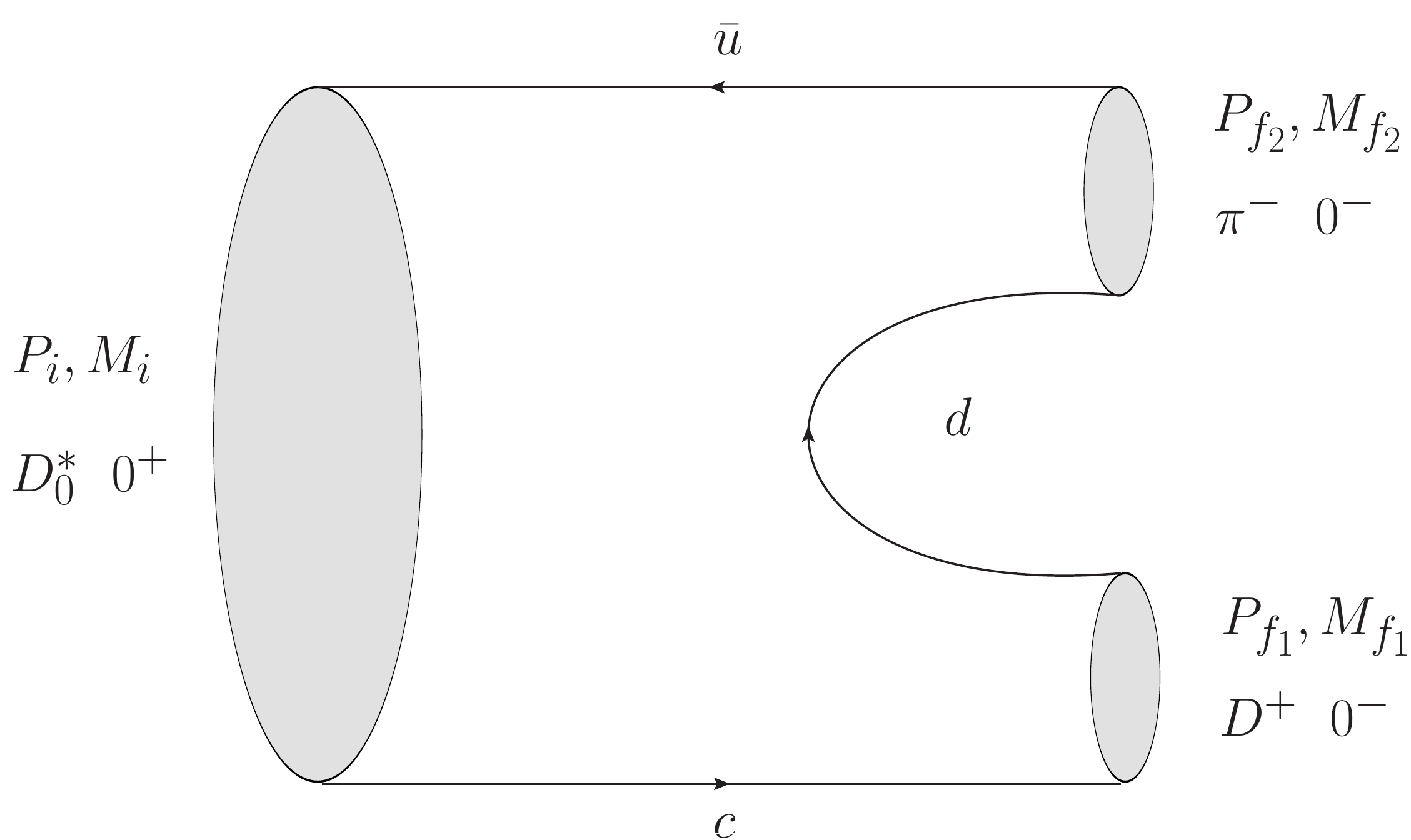}
	\caption[fig1]{Feynman diagram for the decay channel $D_0^*(2400)^0 \to D^+ \pi ^- $.}
	\label{feynman1}
\end{figure}

We take the channel $D_0^*(2400)^0 \to D^+ \pi ^- $ as an example to illustrate the calculation details. The Feynman diagram of this  process is shown in Fig.~\ref{feynman1}.

\begin{figure}[htb]
	\centering
	\includegraphics[scale=0.42]{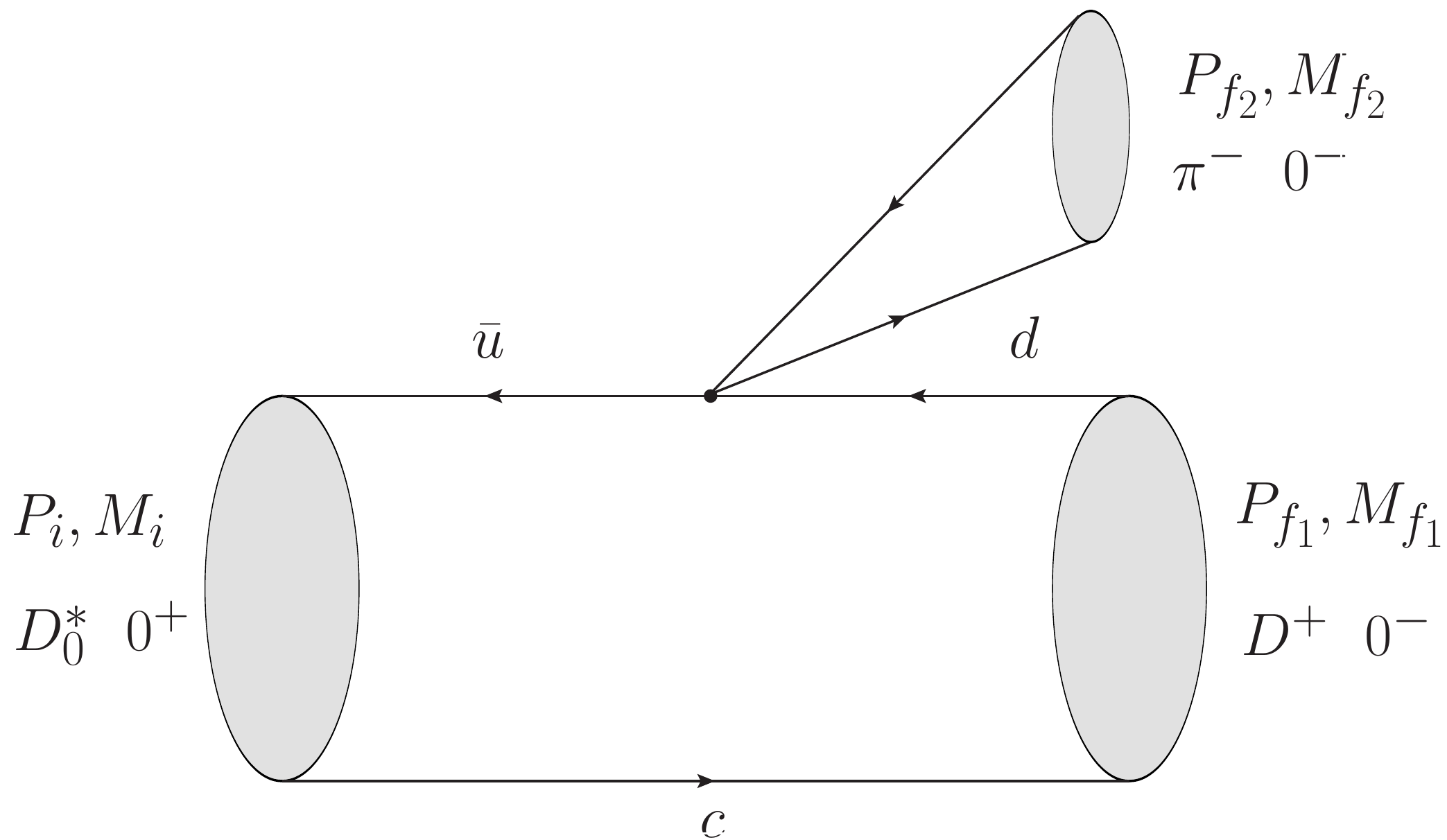}
	\caption[fig1]{Feynman diagram for $D_0^*(2400)^0 \to D^+ \pi ^- $ (with the low-energy approximation).}
	\label{feynman2}
\end{figure}

By using the reduction formula, the transition matrix element can be written as 
\begin{equation}\label{smatrix}
\begin{split}
T=&\langle D^+(P_{f1})\pi ^-(P_{f2})\left| D_0^*(P_i)\right. \rangle \\
=&\int \ud ^4x \ue ^{ \ui P_{f2}\cdot x} (M_{f2}^2 -P_{f2}^2)\langle D^+(P_{f1})\left| \phi _{\pi}(x)\right| D_0^*(P_i) \rangle ,
\end{split}
\end{equation}
where, $\phi_{\pi}$ is the light pseudo-scalar meson field. By using the PCAC relation, the field can be expressed as  \cite{WangGL2005} 
\begin{equation}\label{axialvector}
\phi_{\pi}(x)=\frac{1}{M_{f2}^2 f_{\pi}}\partial ^{\mu}(\overline{u}\gamma _{\mu} \gamma_{5}d),
\end{equation}
where $M_{f2}$ is the mass of $\pi$, and $f_{\pi}$ is its decay constant.

Inserting Eq.~(\ref{axialvector}) into Eq.~(\ref{smatrix}), the transition matrix can be written as 
\begin{equation}
\begin{split}
T=&\frac{M_{f2}^2-P_{f2}^2}{M_{f2}^2 f_{\pi}}\int \ud ^4 x \ue ^{iP_{f2}\cdot x} \langle D^+(P_{f1}) \left| \partial ^{\mu}(\overline{u}\gamma _{\mu} \gamma_{5}d) \right| D_0^*(P_i)\rangle   \\
=&\frac{-\ui P_{f2}^{\mu}(M_{f2}^2-P_{f2}^2)}{M_{f2}f_{\pi}}\int \ud ^4 x \ue ^{iP_{f2}\cdot x} \langle D^+(P_{f1})\left| \overline{u}\gamma _{\mu} \gamma_{5}d \right| D_0^*(P_i)\rangle .
\end{split}
\end{equation}
According to the low energy theorem \cite{WangGL2005}, the momentum of the light meson is much smaller than its mass and can be ignored. Then the Feynman diagram turns to Fig.~\ref{feynman2} and the amplitude can be written as 
\begin{equation}\label{smatrix2}
\begin{split}
T \approx &-\ui \frac{P_{f2}^{\mu}}{f_{\pi}}\int \ud ^4 x \ue ^{\ui P_{f2}\cdot x} \langle D^+(P_{f1})\left| \overline{u}\gamma _{\mu} \gamma_{5}d \right| D_0^*(P_i)\rangle  \\
=&-\ui \frac{P_{f2}^{\mu}}{f_{\pi}} (2\pi)^4 \delta ^4 (P_i -P_{f1}-P_{f2}) \langle D^+(P_{f1})\left| \overline{u}\gamma _{\mu} \gamma_{5}d \right| D_0^*(P_i)\rangle .
\end{split}
\end{equation}

Besides using the PCAC rule and low energy theorem, we also use the effective Lagrangian method to get the transition amplitude of this process and the results of these two approaches are consistent. The Lagrangian is introduced by \cite{Zhong2010el,Wangth2017,lsc2018},

\begin{equation}
	\mathcal{L}_{qqP}=\frac{g}{\sqrt{2}f_h} \bar{q}_{i} \gamma^{\xi} \gamma^5 q_j \partial_{\xi}\phi_{ij},
\end{equation}
where
\begin{equation}
\phi_{ij}=\sqrt{2}
	\left[
	\begin{matrix}
	\frac{1}{\sqrt{2}} \pi^0 + \frac{1}{\sqrt{6}}\eta & \pi^+ & K^+ \\
	\pi^-	&	\-\frac{1}{\sqrt{2}}\pi^0 + \frac{1}{\sqrt{6}}\eta	&	K^0	\\
	 K^-	&	K^0	&	-\frac{2}{\sqrt{6}}\eta
	\end{matrix}
	\right]
\end{equation}
is the chiral field of the pseudoscalar meson. The quark-meson coulping constant $g$ is taken to be unity and $f_h$ is the decay constant.

Within Mandelstam formalism \cite{Mandelstam1955}, we can write the hadronic transition amplitude as the overlap integral over the relativistic wave functions of the initial and final mesons \cite{Chang2006} 
\begin{equation}\label{feynmanamp}
\begin{split}
\mathcal{M}=& -\ui \frac{P_{f2}^{\mu}}{f_{\pi}}\langle D^+(P_{f1})\left| \overline{u}\gamma _{\mu} \gamma_{5}d \right| D_0^*(P_i)\rangle \\
=&-\ui \frac{P_{f2}^{\mu}}{f_{\pi}} \int \frac{\ud ^3 q}{(2 \pi)^3} \mathrm{Tr}\left[\overline{\varphi} _{P_{f1}}^{++}({q}_{f1\perp})\frac{\slashed{P}_i}{M_i}\varphi_{P_i}^{++}({q}_{\perp})\gamma _{\mu}\gamma _{5}\right],
\end{split}
\end{equation}
where $q$ and ${q}_{f1}$ are the relative momenta between quark and anti-quark in initial and final meson, respectively. For the initial meson $D^*_0(c\bar u)$, $q=p_c-\frac{m_c}{m_u+m_c}P_i=\frac{m_u}{m_u+m_c}P_i-p_u$, where $m_u$, $m_c$ are the quark masses and $p_u$ and $p_c$ are the quark momenta.  And for the final meson $D^+(c\bar{d})$, due to the conservation law of momentum, its internal relative momentum is related to that of the initial meson by $q_{f1}=q-\frac{m_c}{m_c+m_d}P_{f1}$. Then, only the BS wave functions in the transition amplitude need to be figured out.

The BS equation of two-body bound state can read in momentum space as \cite{BS1951,LiQ2017}
\begin{equation}
	S_1^{-1}\chi_P(q) S_2^{-1} = \ui \int \frac{\ud ^4 k}{(2\pi)^4}I(P;q,k)\chi_P(k) ,
\end{equation}
where $\chi_P(q)$ is the four-dimensional BS wave function; $I(P;q,k)$ is the interaction kernel; $S_1$ and $S_2$ are the propagators for the quark and anti-quark respectively.

We follow Salpeter \cite{Salpeter1952} to take the instantaneous approximation $I(P;q,k) \approx I(q_{\perp}-k_{\perp})$
The three-dimensional salpeter wave function $\psi(q_{\perp})$ is defined by 
\begin{equation}
	\psi(q_{\perp}) = \ui \int \frac{\ud q_P}{2\pi}\chi_P(q),\quad \chi_P(q)=S_1(p_1) \int \frac{\ud ^3 k}{(2\pi)^3} I(q_{\perp}-k_{\perp})\psi_P(k_{\perp}) S_2(p_2)
\end{equation}

In this work, we adopt the Cornell potential as the interaction kernel $I(r)$ as follow form \cite{WangGL2004,LiQ2017}
\begin{equation}
	I(r) = V_s(r)+V_0+\gamma_0 \otimes \gamma^0 V_v(r) = \frac{\lambda}{\alpha}(1-\ue^{-\alpha r}) + V_0 - \frac{4}{3}\frac{\alpha _s}{r}\ue ^{-\alpha r},
\end{equation}
where $\lambda$ is the string constant, $\alpha_s(r)$ is the running strong coupling constant and $V_0$ is an adjustable parameter fixed by the meson's mass. In momentum space, the potential can read as 
\begin{equation}
	I(\vec{q})= -\left( \frac{\lambda}{\alpha} +V_0	\right) (2\pi)^3 \delta^3(\vec{q})+\frac{\lambda}{\pi^2}\frac{1}{(\vec{q}^2 + \alpha^2)^2} - \frac{2}{3\pi^2}\frac{\alpha_s(\vec{q})}{(\vec{q}^2)+\alpha^2},
\end{equation}
where the coupling constant $\alpha_s(\vec{q})$ is defined by:
\begin{equation}
	\alpha_s(\vec{q})=\frac{12\pi}{27}\frac{1}{\log(\alpha+\frac{\vec{q}^2}{\Lambda_{QCD}^2})}.
\end{equation}

In the above process, we take the instantaneous approximation in the interaction kernel, where we omit the retardation effect. According to the results of paper \cite{Qiao1996,Qiao1999,Ebert2000}, this effect affects much on the light mesons, but has limited influence on the heavy-flavor mesons, because these mesons have larger mass values. 
In addition, retardation effect mainly affects the mass spectra prediction. When we calculate the decay width, we adjust the $V_0$ to  match the experimental data, which further reduces this effect. The results of our previous work \cite{FuHF2011,LiQ2017} are agree with experimental data very well, so the instantaneous approximation is applicable for heavy-light mesons.

Then, we express the relativistic wave function  of a scalar meson with instantaneous approximation ($P_i\cdot q=0$) as 
\begin{equation}
\varphi_{0^+}(q_{\perp})=M \left[
\frac{\slashed{q}_{\perp}}{M}f_{a1}(q_{\perp})+\frac{\slashed{P}\slashed{q}_{\perp}}{M^2} f_{a2}(q_{\perp})+f_{a3}(q_{\perp})+\frac{\slashed{P}}{M}f_{a4}(q_{\perp})
\right],
\end{equation}
where $f_{ai}$($i=1,2,3,4$) are the functions of $q_{\perp}^2$ and their value can be obtained by solving the full Salpeter equations. It is notable that $\varphi_{0^+}(q_{\perp})$ is a general form for $J^P=0^+$ states and the items containing $q$ are the high order relativistic corrections.

Within BS method, the four wave functions $f_{ai}$ are not independent, they have the following relations \cite{WangGL2009} 
\begin{equation}
\begin{split}
&f_{a3}=\frac{q_{\perp}^2(\omega_1+\omega_2)}{M(m_1 \omega_2+m_2 \omega_1)}f_{a1}, \\
&f_{a4}=\frac{q_{\perp}^2(\omega_1-\omega_2)}{M(m_1\omega_2+m_2\omega_1)}f_{a2},
\end{split}
\end{equation}where $m_1=m_c$, $m_2=m_u$, $\omega_1=\sqrt{m_1^2-{q}_{\perp}^2}$, and $\omega_2=\sqrt{m_2^2-{q}_{\perp}^2}$.

In our calculation, we only keep the positive energy parts $\varphi_{P_i}^{++}({q}_{i\perp})$ of the relativistic wave functions because the negative energy part contributes too small \cite{Wangth2017}.
The positive energy part of the wave function can be written as 
\begin{equation}\label{0+}
\varphi_{0^+}^{++}(q_{\perp})=A_1+A_2\frac{\slashed{P}}{M}+A_3\frac{\slashed{q}_{\perp}}{M}+A_4\frac{\slashed{P}\slashed{q}_{\perp}}{M^2},
\end{equation}
where
\begin{equation}
\begin{split}
A_1=&\;\frac{(\omega_1+\omega_2)q^2_{\perp}}{2(m_1\omega_2+m_2\omega_1)}\left(
f_{a1}+\frac{m_1+m_2}{\omega _1+\omega_2}f_{a2}
\right), \\
A_2=&\ \frac{(m_1-m_2)q^2_{\perp}}{2(m_1\omega_2+m_2\omega_1)}\left(
f_{a1}+\frac{m_1+m_2}{\omega _1+\omega_2}f_{a2}
\right), \\
&A_3=\frac{M}{2}\left(
f_{a1}+\frac{m_1+m_2}{\omega_1+\omega_2}f_{a2}
\right), \\
&A_4=\frac{M}{2}\left(
\frac{\omega_1+\omega_2}{m_1+m_2}f_{a1}+f_{a2}
\right).
\end{split}
\end{equation}
To calculate the values of wave functions, we should determine the parameters' values in the interaction kernel. We try to fix $V_0$ by the mass of the ground state. In this case, the theoretical mass of $D_J^*(3000)$ is much less than the present experimental data. Thus, we adjust $V_0$ to make its mass value be equal to the experimental data, then get the wave functions.
In this work, besides the wave function for $0^+$ state, we also need the wave functions of $0^-$, $1^-$, $1^+$, etc., which are presented in the appendix.

After finishing the integral, we can get the amplitude of $0^+ \to 0^-0^-$ as follow 
\begin{equation}
\begin{split}
\mathcal{M}_{(0^+ \to 0^- 0^-)}=-\ui \frac{P_{f2}^{\mu}}{f_{\pi}}
(P_{\mu} n_1 + P_{f1 \mu} n_2) ,
\end{split}
\end{equation}
where $n_1$ and $n_2$ are the form factors. They are the overlap integral over the wave functions of the initial and final states.

If the final light meson is $\eta$ or $\eta '$, the $\eta-\eta '$ mixing should be considered 
\begin{equation}
\left(
\begin{array}{c}
\eta \\
\eta '
\end{array}
\right)
=
\left(
\begin{array}{cc}
\cos \theta_P & -\sin \theta_P \\
\sin \theta_P & \cos \theta_P
\end{array}
\right)
\left(
\begin{array}{c}
\eta_8 \\
\eta_1
\end{array}
\right),
\end{equation}
where $\eta_1= (u\bar{u}+d\bar{d}+s\bar{s})/{\sqrt{3}}$ and $\eta_8= {(u\bar{u}+d\bar{d}-2s\bar{s})}/{\sqrt{6}}$, we choose the mixing angle $\theta_P=-11.4^ \circ$ \cite{PDG2016}. Then, we get the transition amplitude with an extra coefficient after considering the mixing 
\begin{equation}
\begin{split}
\mathcal{M}(\eta)=- \ui P_{f2}^{\mu} M^2_{\eta} \left(
\frac{\cos \theta_P}{\sqrt{6}f_{\eta_8}M^2_{\eta_8}}-\frac{\sin \theta_P}{\sqrt{3}f_{\eta_1}M^2_{\eta_1}}
\right) \langle D^0(P_{f1})\left| \overline{u}\gamma _{\mu} \gamma_{5}u \right| D_0^*(P_i)\rangle, \\
\mathcal{M}(\eta ')=- \ui P_{f2}^{\mu} M^2_{\eta '}\left(
\frac{\sin \theta_P}{\sqrt{6}f_{\eta_8}M^2_{\eta_8}}+\frac{\cos \theta_P}{\sqrt{3}f_{\eta_1}M^2_{\eta_1}}
\right) \langle D^0(P_{f1})\left| \overline{u}\gamma _{\mu} \gamma_{5}u \right| D_0^*(P_i)\rangle.
\end{split}
\end{equation}

In the case when heavy-light $1^+$ state is involved, if we use the $S$-$L$ coupling, the $^3P_1$ and $^1P_1$ states cannot describe the physical states. Within the heavy quark limit($m_Q\to \infty$), its spin decouples and the properties of the heavy-light $1^+$ state are determined by those of the light quarks. So $j$-$j$ coupling should be used instead. The orbital angular momentum $\vec{L}$ couples with the light quark spin $\vec{s}_q$, which is $\vec{j}_l=\vec{L}+\vec{s}_q$. Then $1^+$ state can be grouped into a doublet by the total angular momentum of the light quark($|j_l=1/2\rangle$ and $|j_l=3/2\rangle$). The relation between the two descriptions are \cite{Matsuki2010mixingangle,Barnes2005highercharmonia}
\begin{equation}
\left(
\begin{array}{c}
|J^P=1^+,j_l={3}/{2} \rangle \\
|J^P=1^+,j_l={1}/{2} \rangle \\
\end{array}
\right)
=
\left(
\begin{array}{cc}
\cos \theta  & \sin \theta \\
-\sin \theta & \cos \theta \\
\end{array}
\right)
\left(
\begin{array}{c}
|{^1}P_1 \rangle \\
|{^3}P_1 \rangle \\
\end{array}
\right).
\end{equation}
In our method, we solve the Salpeter equations for $^3P_1$ and $^1P_1$ states individually, and use these mixing relations to calculate the contributions of two physical $1^+$ states.
We list some mixing states related to our work 
\begin{equation}
\left(
\begin{array}{c}
D_1(2420) \\
D_1(2430) \\
\end{array}
\right)
=
\left(
\begin{array}{cc}
\cos \theta  & \sin \theta \\
-\sin \theta & \cos \theta \\
\end{array}
\right)
\left(
\begin{array}{c}
D (1{^1}P_1) \\
D (1{^3}P_1) \\
\end{array}
\right),
\end{equation}
\begin{equation}
\left(
\begin{array}{c}
D_{s1}(2536) \\
D_{s1}(2460) \\
\end{array}
\right)
=
\left(
\begin{array}{cc}
\cos \theta & \sin \theta \\
-\sin \theta & \cos \theta \\
\end{array}
\right)
\left(
\begin{array}{c}
D_s (1 {^1}P_1) \\
D_s (1 {^3}P_1 )\\
\end{array}
\right).
\end{equation}
In our calculation, for these doublets, we choose the ideal mixing angle $\theta=35.3^{\circ}$ in the heavy quark limit.

For the $^3P_1(1^{++})$ and $^1P_1(1^{+-})$ states, the corresponding hadronic transition amplitudes are 
\begin{equation}
\begin{split}
\mathcal{M}_{(0^+\to 1^{++} 0^-)}= \frac{-\ui}{f_{\pi}} \varepsilon_{1\mu}P^{\mu} t_1, \\
\mathcal{M}_{(0^+\to 1^{+-} 0^-)}= \frac{-\ui}{f_{\pi}} \varepsilon_{1\mu}P^{\mu} t_2,
\end{split}
\end{equation}
where $\varepsilon$ is the polarization vector of the $1^+$ state; $t_1$ and  $t_2$ are the form factors. Then, the form factors of the physical states are 
\begin{equation}
	\begin{split}
    t_{D_1(2420),D_{s1}(2536)}= t_2 \cos \theta + t_1 \sin \theta, \\
	t_{D_1(2430),D_{s1}(2460)}=-t_2 \sin \theta + t_1 \cos \theta.
	\end{split}
\end{equation}

The PCAC rule can only be applied to light pseudo-scalar mesons and it is not valid for light vector meson.
If  $\rho$ or $\omega$ meson appears in the final states, we choose the effective Lagrangian method to calculate the transition amplitude. The Lagrangian of quark-meson coupling can be expressed as \cite{Zhong2010el,Wangth2017,lsc2018} 
\begin{equation}\label{efl}
	\mathcal{L}_{qqV}= \bar{q}_i(a \gamma_{\mu}+\frac{\ui b}{2 M_{P_{f2}}}\sigma_{\mu \nu}P_{f2}^\nu )V_{ij}^{\mu}q_j,
\end{equation}
where $V^{\mu}_{ij}$ is the field of the light vector meson; $q_i$ and $\bar q_j$ are its constitute quarks. And we choose the parameters $a=-3$ and $b=2$ which represent the vector and tensor coupling strength \cite{Wangth2017}, respectively. Then we use Eq.~(\ref{efl}) to derive the light-vector meson's vertex and  get the transition amplitude 
\begin{equation}\label{feynmanampefl}
\begin{split}
\mathcal{M}=-\ui \int \frac{\ud ^3 q}{(2 \pi)^3} \mathrm{Tr}\left[\overline{\varphi} _{P_{f1}}^{++}({q}_{f1\perp})\frac{\slashed{P}_i}{M_i}\varphi_{P_i}^{++}({q}_{\perp})(a \gamma_{\mu}+\frac{\ui b}{2 M_{f2}}\sigma_{\mu \nu}P_{f2}^\nu )\varepsilon_{2}^{\mu}\right] .
\end{split}
\end{equation}

After finishing the trace and integral, the transition amplitudes can be expressed as 
\begin{equation}
	\begin{split}
	\mathcal{M}_{(0^+ \to 1^- 1^{-})}=\varepsilon_{1\mu} \varepsilon_2^{\mu} t_1
	+\varepsilon_{1\mu}P^{\mu} \varepsilon_{2\nu} P^{\nu}  t_2,
	\end{split}
\end{equation}
where $\varepsilon_{1\mu}$ and $\varepsilon_{2\nu}$ are the polarization vectors of final heavy vector meson and the light vector meson, respectively; $t_1$, $t_2$ and $t_3$ are the form factors. 

Then, the two-body decay width can be expressed as 
\begin{equation}
\begin{split}
\Gamma=\frac{1}{8 \pi} \frac{\vert \vec{P}_{f1} \vert}{M^2_i}|\mathcal{M}|^2 ,
\end{split}
\end{equation}
where $\vec{P}_{f1}$ is the three-dimensional momentum of the final charmed meson 
\begin{equation}
	\vert \vec{P}_{f1} \vert = \sqrt{\left(\frac{M_i^2+M_{f1}^2-M_{f2}^2}{2M_i}\right)^2-M_{f1}^2}  .
\end{equation}

\section{Results and Discussions}

In this paper, the masses of constituent quarks that we adopt are listed as follows:  $m_u=0.305\GeV,\ m_d=0.311\GeV,\ m_s=0.50\GeV,\ {\rm and}~  m_c=1.62\GeV$ \cite{Wangth2013}. Other parameters are $\alpha=0.060\GeV$, $\lambda=0.210\GeV^2$, $\Lambda_{QCD}=0.270 \GeV$, $f_{\pi}=0.1304\GeV$, $f_K=0.1562\GeV$ \cite{PDG2016}, $f_{\eta_1}=1.07f_{\pi}$, $f_{\eta_8}=1.26f_{\pi}$, $M_{\eta_1}=0.923\GeV$, and $M_{\eta_8}=0.604\GeV$ \cite{Wangth2017}. The masses of other involved mesons are shown in Table \ref{canshu}. 

\begin{table}[htb]
	\renewcommand\arraystretch{1.01}
	\caption[canshu]{Masses of involved mesons ($ \mathrm{GeV} $) \cite{PDG2016}. }
	\label{canshu}
	\vspace{0.5em}\centering
	\begin{tabular}{cccc}
		\toprule[1.5pt]
		{\quad $ m_{D_0^*(2400)^0}=2.318  $\quad } & $ m_{D_0^*(2400)^+}=2.351  $	& \quad {$ m_{D_J^*(3000)^{(0,+)}}=3.008 $} &  \quad {$ m_{D_{s}^+}=1.968 $} \quad \\
		$ m_{D_1(2420)^0}=2.421  $ & $ m_{D_1(2420)^+}=2.423 $  & \quad $ m_{D_1(2430)^{(0,+)}} =2.427  $ & \quad 	{$ m_{D_s^{*+}}=2.112 $}\\
		\bottomrule[1.5pt]
	\end{tabular}
\end{table}

We first calculate the the decay widths of the $1P$ states. It only have two OZI-allowed decay channels and the results are presented in Table \ref{d2400_0}. In the case of $D_0^*(2400)^0$, the decay width of $D^+\pi^-$ is almost twice as that of $D^0\pi^0$. Because there is a factor $1/\sqrt{2}$ in the constitute quarks of $\pi^0$. Other decays that involve $\rho^0$ and $\omega^0$ have similar relation too.

\begin{table}[htb]
	\renewcommand\arraystretch{0.90}
	\caption[tab1]{$D_0^*(2400)^{0,+}$ strong decay widths (MeV). Ref. \cite{zhaoqiang2008} adopts Chiral Quark Model, Ref. \cite{Close2005} adopts the $^3P_0$ Model and Ref. \cite{Godfrey2005} adopts the Pseudoscalar Emission Model.}
	\label{d2400_0}
	\vspace{0.5em}\centering
	\begin{tabular}{ccccccc}
		\toprule[1.5pt]
		\multicolumn{2}{c}{Chanel} &  \quad Ours\quad & \quad Ref. \cite{zhaoqiang2008} \quad & \quad Ref. \cite{Close2005} \quad & \quad Ref. \cite{Godfrey2005} & \quad Exp. \cite{PDG2016}	\\
		\midrule[1pt]
		\multirow{2}{*}{$D_0^*(2400)^0\to $} & $D^+ \pi^-$ &  \quad 151.5 \quad  &  \multirow{2}{*}{266} & \multirow{2}{*}{283} & \multirow{2}{*}{277} & \multirow{2}{*}{$267\pm 40$}	\\
		{ } & $ D^0 \pi^0$ &\quad 74.8 \quad&  { } & & & 	\\
		\multirow{2}{*}{$D_0^*(2400)^+\to $} & $ D^+ \pi^0$ & \quad 81.6 \quad& \multirow{2}{*}{$\Box$} & \multirow{2}{*}{$\Box$} &\multirow{2}{*}{$\Box$} & \multirow{2}{*}{$230\pm17$}	\\
		{ }& $ D^0 \pi^+$ & \quad 164.3 \quad &  { }& & & 	\\
		\bottomrule[1.5pt]
	\end{tabular}
\end{table}

\begin{figure}[htb]
	\centering
	\subfigure[$\Gamma_{  D_0^*(2400)^{(0,+)}}$ versus the mass]{
		\label{width_mass_2400} 
		\includegraphics[width=0.45\textwidth]{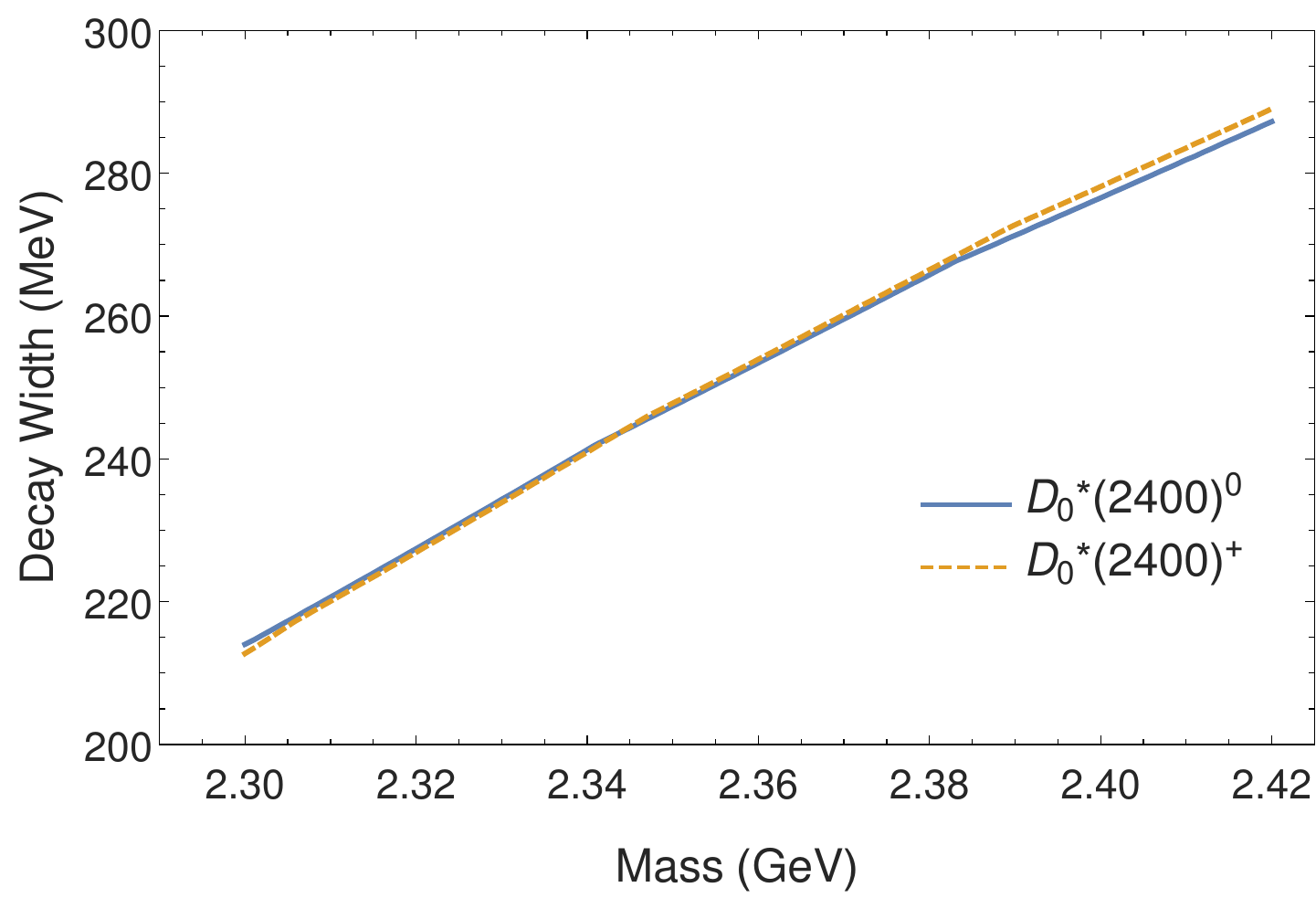}}
	\hspace{0.2cm}
	\subfigure[$\Gamma_{ D_J^*(3000)^{(0,+)}}$ versus the mass]{
		\label{width_mass_3000} 
		\includegraphics[width=0.45\textwidth]{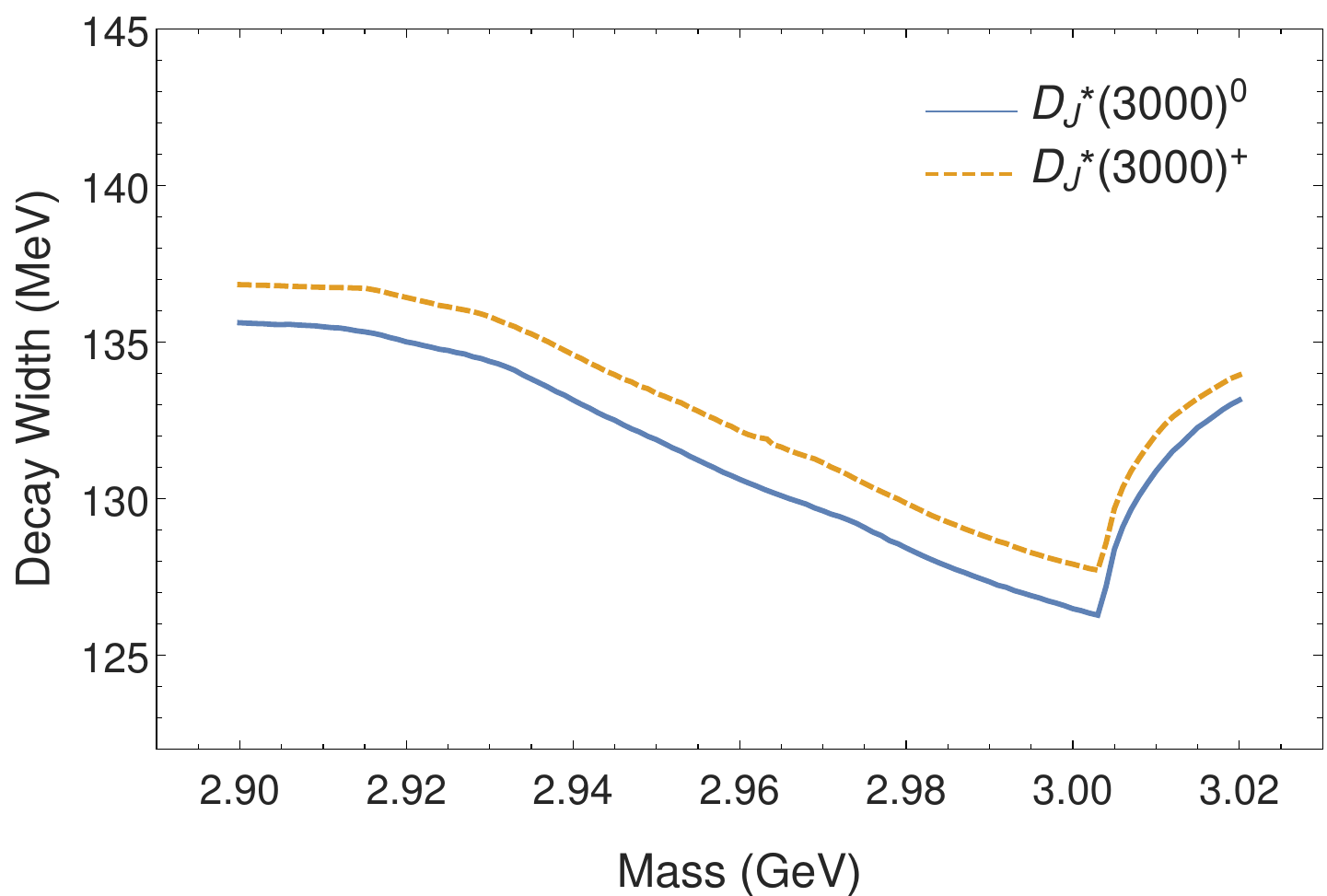}}
	\label{width_mass}
	\caption{Total decay widths of $D_0^*(2400)$ and $D_J^*(3000)$ change with the masses.}
\end{figure}

The total decay width of $D_0^*(2400)^+$ is larger than that of $D_0^*(2400)^0$ in our calculation, which are 245.9 MeV and 226.3 MeV, respectively. According to the present experimental data, the charged $D_0^*(2400)^{+}$ is heavier than the neutral $D_0^*(2400)^{0}$. The different phase spaces may result in this discrepancy. We also notice that the estimated decay widths are sensitive to the mass of the initial meson. Considering the experimental mass values have errors ($m_{D_0^*(2400)^0}=2318\pm 29 \ \mathrm{MeV},\ m_{D_0^*(2400)^{\pm}}=2351\pm \ 7 \mathrm{MeV}$ \cite{PDG2016}) and these experimental masses value have divergence with different theoretical predictions\cite{Godfrey1985,Pierro2001,Ebert2009,Sun2013,Godfrey2016}, we give the two-body decay width changing along with the initial meson mass from 2300 MeV to 2420 MeV, which is shown in Fig.~\ref{width_mass_2400}. The neutral one's total decay width changes from 214.0 to 287.2 MeV, and the charged one's is from 212.7 to 289.0 MeV . We believe that these OZI-allowed decays happen around the mass threshold, which results in such sensitivity of decay width to the initial $1P$ state mass.

In Table \ref{d2400_0}, we also list the results from other models  \cite{Godfrey2005,zhaoqiang2008,Close2005} as well as the experimental results for comparison. According to Table \ref{d2400_0} and Fig.~\ref{width_mass_2400}, we conclude that our results of the $1P$ states are consistent with experimental data, which means we can apply the same method to study the $2P$ states.

\begin{table}[htb!]
	\renewcommand\arraystretch{0.99}
	\caption[D3000duibi]{Two-body strong decay widths (MeV) of $ D_J^*(3000)^0 $ as the $2P(0^+)$ state. ``-'' means the channel is forbidden, ``$\Box$'' means the channel is not included by this method.  Ref. \cite{YuGL2015} uses $^3P_0 $ Model; Ref. \cite{Sun2013} uses QPC Model; Ref. \cite{Godfrey2016} uses Relativistic quark model and Ref. \cite{Gupta2018}. uses effective Lagrangian approach. }\label{D3000results}
	\vspace{0.3em}\centering
	\begin{tabular}{llccccc}
		\toprule[1.5pt]
		{Chanel }\qquad \quad \quad  & Final States \quad 	&\quad 	Ours \quad   & \quad Ref. \cite{YuGL2015}\ \  & \quad  Ref. \cite{Sun2013} &\quad
		Ref. \cite{Godfrey2016} &\quad Ref. \cite{Gupta2018}\\
		\midrule[1pt]
		\multirow{2}{*}{ $  D(^1S_0) \pi $ }  & $ D^+ \pi^- $ & 11.6 & 23.94 & \multirow{2}{*}{ 49 } & \multirow{2}{*}{25.4} & 66.2 \\
		&	$ D^0 \pi^0 $	&	6.1	& {11.97}	& { } & { }	\vspace{0.3em} & 33.3\\
		\multirow{2}{*}{$ D(2 ^1S_0)\pi$ }& $D(2550)^+ \pi^-$& 6.9 & \multirow{2}{*}{$\Box$} & \multirow{2}{*}{  $\Box$ } & \multirow{2}{*}{ 18.6} & \multirow{2}{*}{  $\Box$ }\\
		& $D(2550)^0 \pi^0 $ & 3.3 & { } & 	{ }	& { } \vspace{0.6em}	&  \\
		{$  D \eta $}  & $ D^0 \eta^0 $    & 	0.51	& 4.26	& 8.8	& 1.53	& 10.8	\\
		{$  D \eta ' $}& $ D^0 \eta '^0$   &	6.0	& 1.07	& 2.7	& 4.94 &	{ $\Box$ } \\
		{$  D_s K$}	& $ D_s^+ K^- $ & \~{}$10^{-3}$  & 2.85 & 6.6 & 0.76 & 54.2 \vspace{0.6em}  \\
		\multirow{2}{*}{$  D_1(2420)\pi$	}& $ D_1(2420)^0 \pi^0 $ & 18.7	& 26.20 & \multirow{2}{*}{38} & \multirow{2}{*}{96.1$(^1P_1)$} & \multirow{2}{*}{  $\Box$ }\\
		&  $ D_1(2420)^+ \pi^- $ &  36.8	& $\Box$ & {} & 	&  \\
		{$  D_1(2420) \eta$}& $ D_1(2420)^0 \eta^0 $ & 0.85 & 1.37 & 1.1  & $\Box$ & $\Box$ \\
		\multirow{2}{*}{$  D_1(2430) \pi$}& $ D_1(2430)^0 \pi^0 $ & 2.1 & 6.69 & \multirow{2}{*}{30} & \multirow{2}{*}{  $\Box$ }	& \multirow{2}{*}{  $\Box$ } \\
		{ } & $ D_1(2430)^+ \pi^- $ & 4.1 & { $\Box$ } & {  } &  & \\
		{$  D_1(2430) \eta$}& $ D_1(2430)^0 \eta ^0 $ & 0.12 & 0.35 & 0.91 	& $\Box$ & $\Box$ \\ \vspace{0.6em}
		{$  D_{s}(2460) K$}& $ D_{s1}(2460)^+ K^- $ & 1.2 & 12.81 & 1.5 & { $\Box$}   \vspace{0.3em} & $\Box$ \\
		\multirow{2}{*}{$  D^{*} \rho$}& $ D^{*}(2007)^0 \rho ^0 $& 7.0 & 31.60 &  \multirow{2}{*}{41} & \multirow{2}{*}{32} & \multirow{2}{*}{  $\Box$ } \\
		& $D^{*}(2010)^+ \rho ^-$ & 13.3 & 62.01 &{ } \vspace{0.6em}  & \\
		{$  D^{*} \omega$}& $ D^{*}(2007)^0 \omega ^0 $ & 7.5 & 29.91 & 13 & 10.2	& $\Box$ \\
		{$  D_s^{*} K^*$}& $ D_s^{*+} K^*(892)^- $ & 4.1 & 3.06 & 1.0	&$\Box$	& $\Box$ \\
		{$  D_s(2536) K^-$}& $D_{s1}(2536)^+ K^-$ & - & 6.40  & - & - \vspace{0.6em} & - \\
		\midrule[1pt]
		\multicolumn{2}{c}{Total} & 130.2 & 224.5 & 193.6 & 189.5 & 164.5\\
		\multicolumn{2}{c}{Experimental value} & \multicolumn{4}{c}{$110.5\pm 11.5$} & \\
		\bottomrule[1.5pt]
	\end{tabular}
\end{table}

\begin{figure}[htbp]
	\centering
	\subfigure[$ 0^+ $($1P$) state $ D_0^*(2400)^0$]{
		\label{bse1p0+} 
		\includegraphics[width=0.45\textwidth]{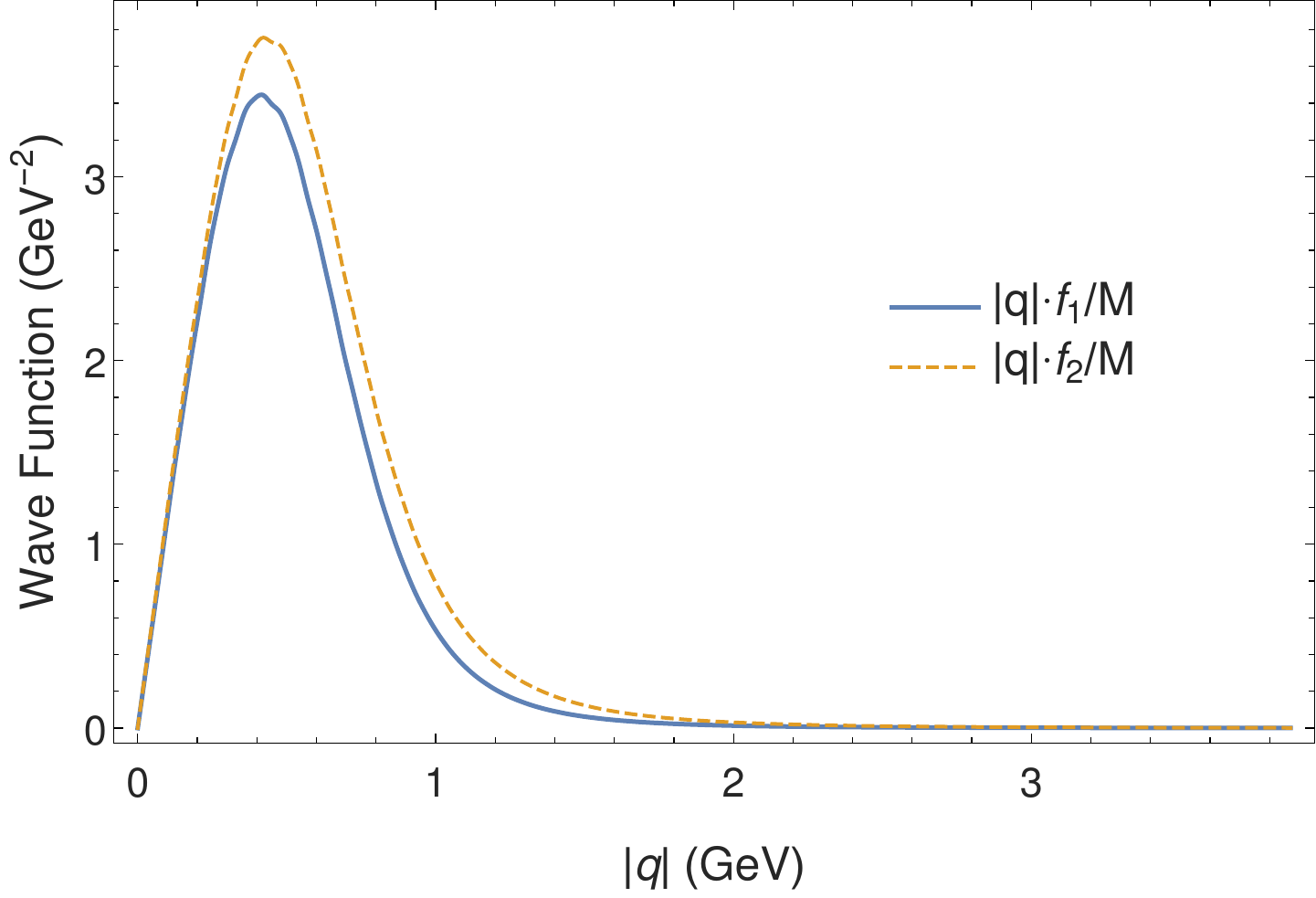}}
	\hspace{0.2cm}
	\subfigure[$ 0^+ $($ 2P $) state $ D_J^*(3000)^0 $]{
		\label{bse2p0+} 
		\includegraphics[width=0.45\textwidth]{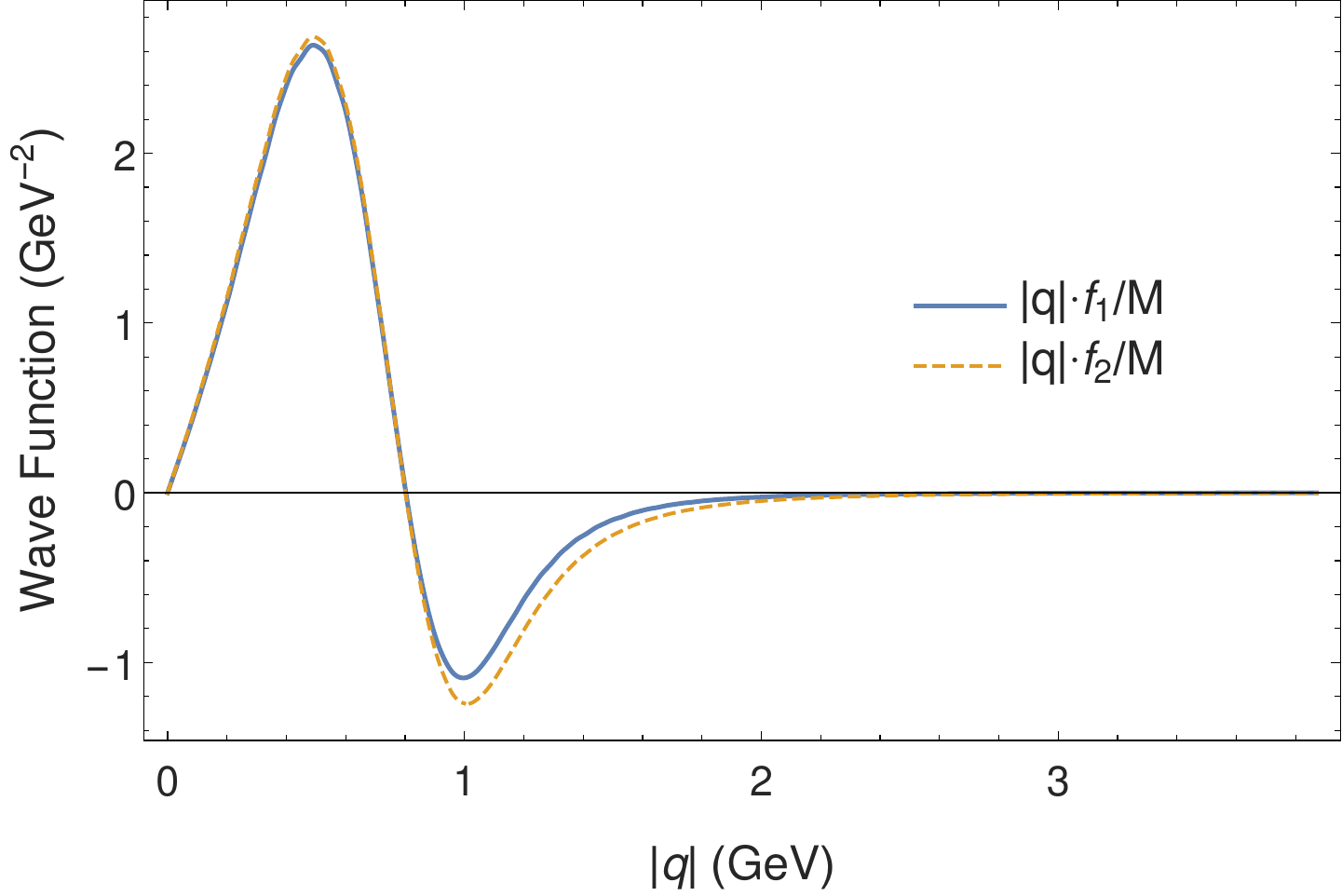}}
	
	\subfigure[$D_J^*(3000)^0 $ and $D_1(2420)$($^3P_1 $ state)]{
		\label{bse1p1++} 
		\includegraphics[width=0.45\textwidth]{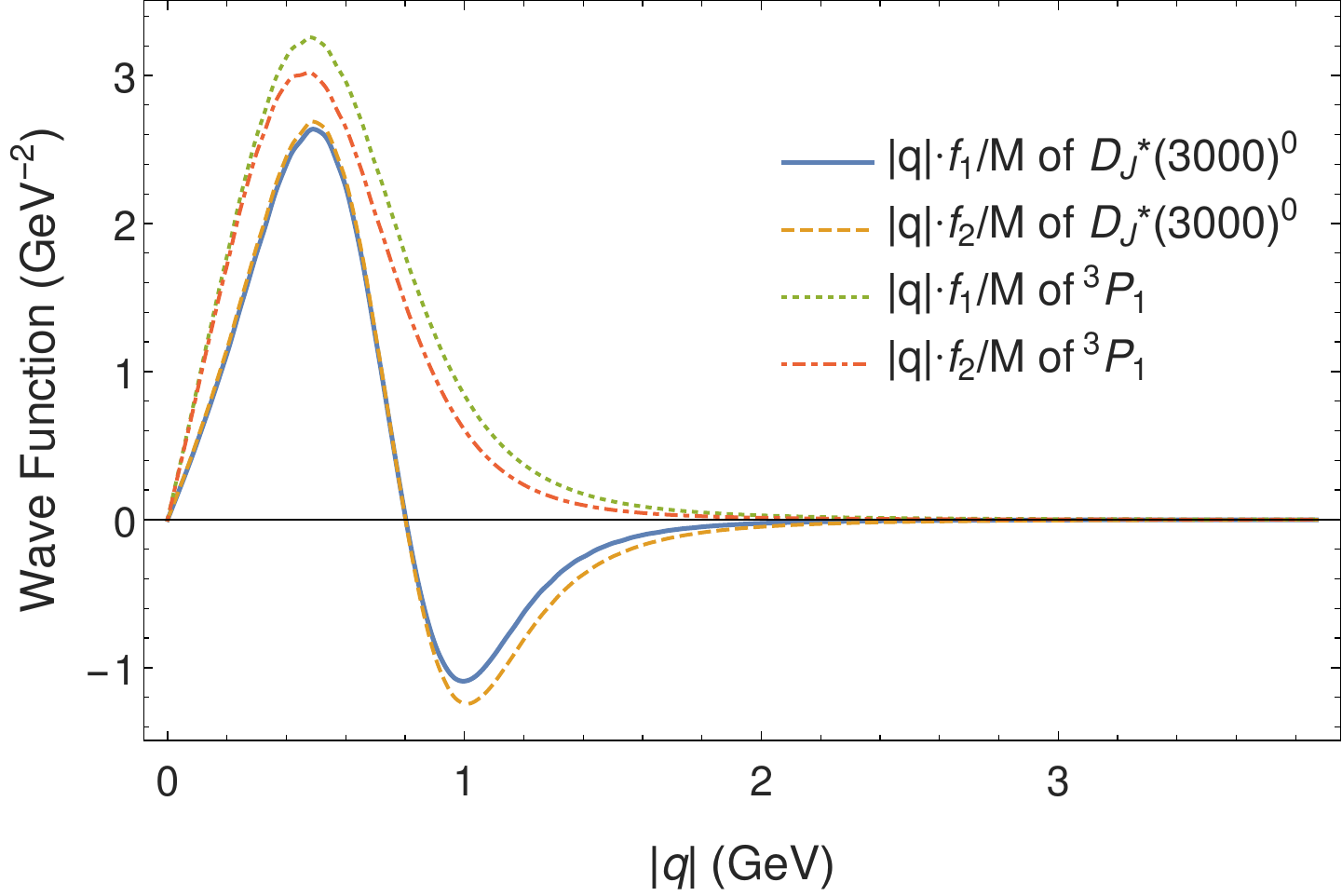}}
	\hspace{0.2cm}
	\subfigure[$D_J^*(3000)^0 $ and $D_1(2420)$($^1P_1$ state)]{
		\label{bse1p1+-} 
		\includegraphics[width=0.45\textwidth]{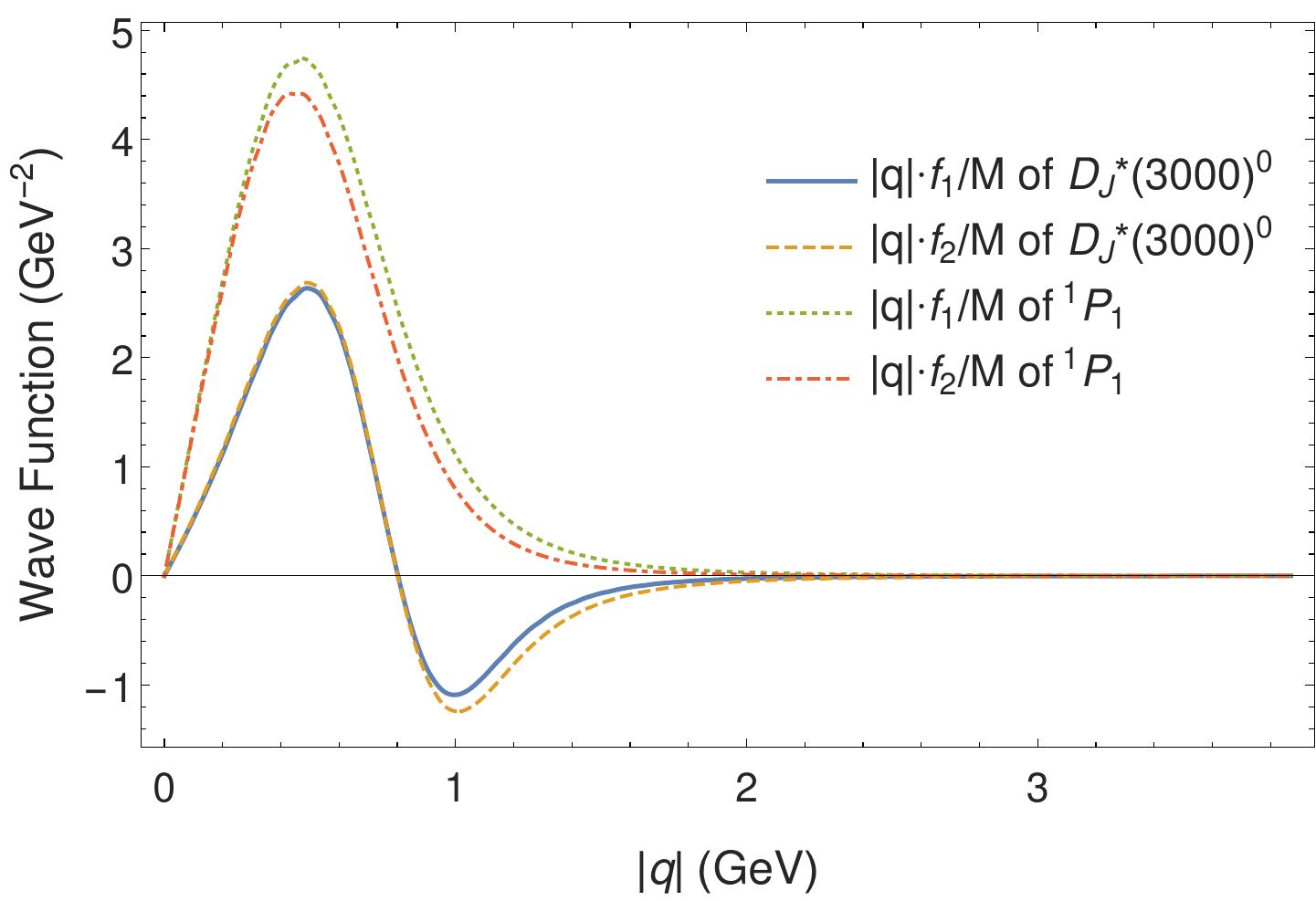}}
	
	\subfigure[$D_J^*(3000)^0 $ and $D^0$($^1S_0$ state)]{
		\label{bse1s0-} 
		\includegraphics[width=0.45\textwidth]{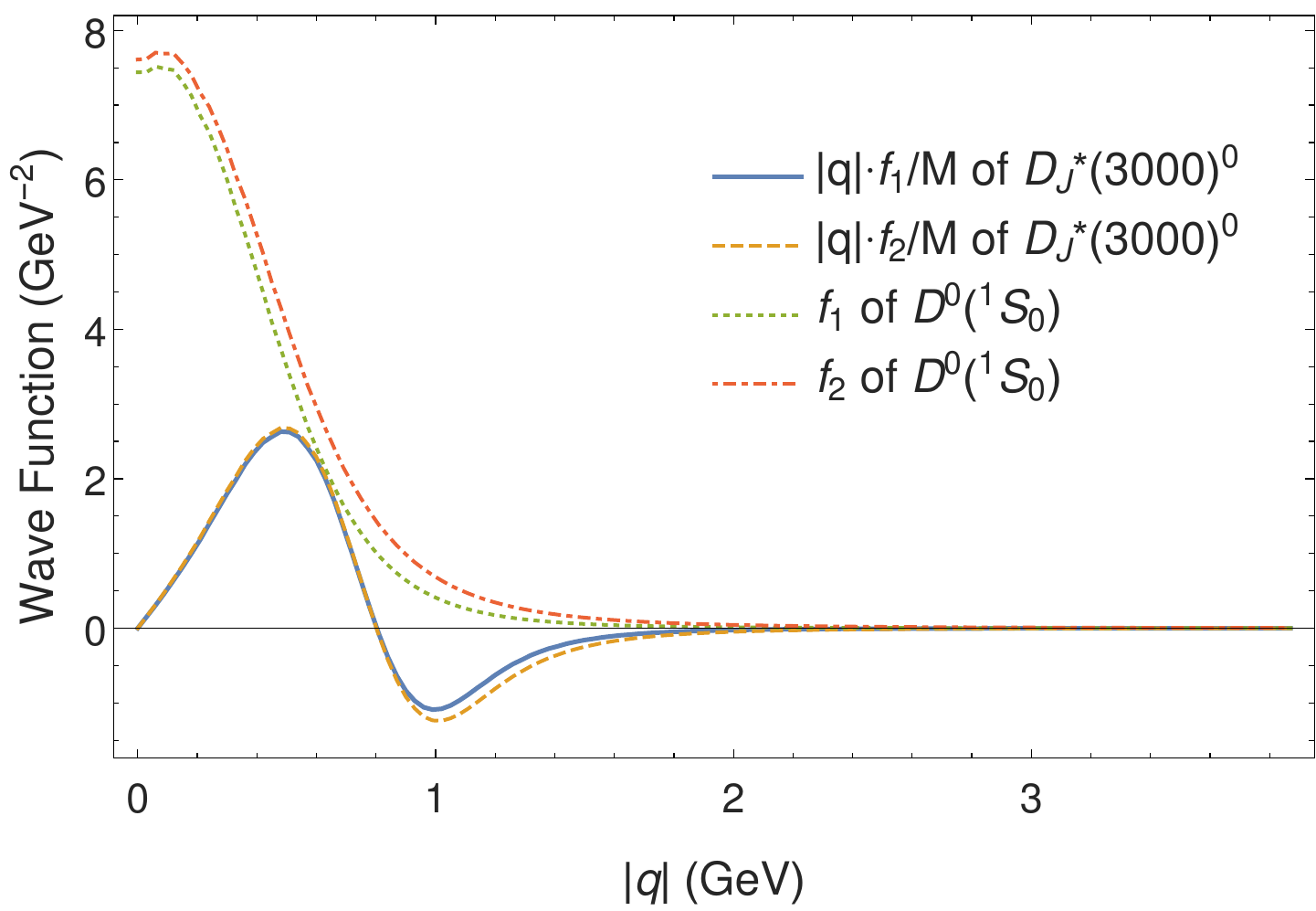}}
	\hspace{0.2cm}
	\subfigure[$D_J^*(3000)^0 $ and $D^0$($2^1S_0$ state)]{
		\label{bse2s0-} 
		\includegraphics[width=0.45\textwidth]{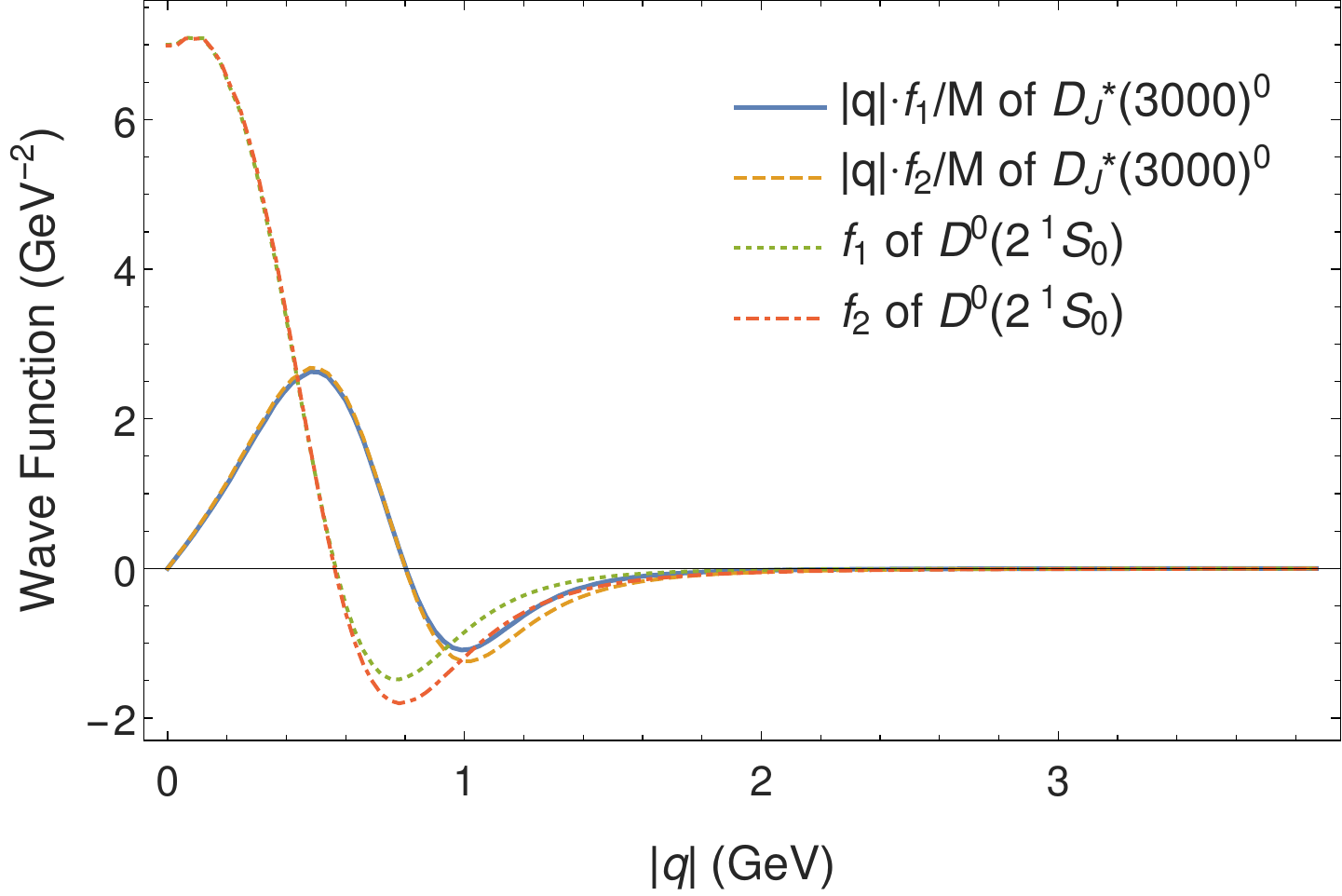}}
	\caption{Several examples of wave functions for some states}
	\label{bseplot} 
\end{figure}

Under the assumption of $0^+$($2P$) state, the results of ours and other models are shown in Table \ref{D3000results}. The total width of our calculation is 130.2 MeV, which is smaller than the results of other models and close to the upper limit of experimental value. Though the $2P$ state $D_J^*(3000)^0$ has larger phase space and more decay channels than those of $1P$ state $D_0^*(2400)^0$, why we get a narrower full width? The reason is the different structures of wave functions. The numerical values of the wave functions $f_{a1}$ and $f_{a2}$ for $1P$ state as the function of internal momentum $|\vec{q}|$ are all positive (Fig.~\ref{bse1p0+}), while the wave functions of the $2P$ state have a node (Fig.~\ref{bse2p0+}). The wave function values after the node become negative and it makes contrary contribution to the positive part, which will cause cancellation in the overlap integral between these two parts. The node structure reduces the decay width of the $2P$ state.

From our results, the channels of $D \pi,\ D_1(2420)\pi ,\ D\rho$ give large contribution to the full decay width, and $D_1(2420)\pi$ channel is dominant.
As shown in Eq.~(\ref{feynmanamp}), large transition amplitude means that these three channels have large overlap integrals. For example, we draw the wave functions of final state $D_1(2420)(^3P_1 \& ^1P_1)$ and initial $D_J^*(3000)$ in Fig.~\ref{bse1p1++} and \ref{bse1p1+-}. When the recoil momentum of the final meson is small (ignoring the difference between internal momenta of the initial and final states here), the peak values of the initial and final wave functions are coincident. Thus, we obtain large overlap integral values of the wave functions before the $2P$ node. However, the part after the node gives small cancellation since the corresponding values of the $1P$ wave function are small at this time. As a result, we obtain a large decay width of $D_1(2420)\pi$ channel.

Other examples of $D(1^1S_0)$ and $D(2^1S_0)$ are shown in Fig.~\ref{bse1s0-} and \ref{bse2s0-}. For the channel of $D(1^1S_0)\pi$, when compared with $D_1(2420) \pi$, the part after the node gives negative contribution.  
Because the peak values of the those wave functions are not coincident, the positive contribution of the part before the node is not dominant. The negative part after the node changes the sign of the overlap integral, which cause the width of $D(1^1S_0)\pi$ is narrower than that of $D_1(2420)\pi$. 
For the $D(2^1S_0)\pi$ channel, both wave functions have node structure. Compared with $D(1^1S_0)\pi$ channel, the contribution from the part before the $2P$ node is obviously smaller for $D(2^1S_0)\pi$. But the contribution after the $2P$ node becomes positive again because both wave functions ($2S$ and $2P$ states) are negative at this time. Therefore, although the phase space is narrow, the decay width of $D(2^1S_0)\pi$ channel is not very small.

It has been mentioned in Sec.~\uppercase\expandafter{\romannumeral1} that the relativistic corrections are large for the excited states. We can explain this argument according to the figures of the wave functions. If contribution from large internal momentum $q$ is significant, we can conclude that relativistic corrections are considerable. In Fig.~\ref{bse1s0-}, for the $1S$ state, the peak value of the wave function appear in the region of small $q$, while for the $1P$ state, in Fig.~\ref{bse1p1++} and \ref{bse1p1+-}, the peak values appear in the region of middle $q$. This means the $1P$ state has larger relativistic correction than that of the $1S$ ground state. When comparing Fig.~\ref{bse1s0-} with \ref{bse2s0-}, wave functions after the node give sizable contribution, which happens in large $q$ region. This means higher excited states have larger relativistic corrections. So we conclude that a relativistic model is needed to deal with the excited state problem.

In our study, we also calculate the decay widths of $ D_J^*(3000)^+ $, shown in Table \ref{D3000plusresults}. All channels are similar to $D_J^*(3000)^0$, and the full width is 131.3 MeV.

\begin{table}[htb!]
	\renewcommand\arraystretch{1.0}
	\caption[D3000duibi]{Two-body strong decay widths (MeV) of $ D_J^*(3000)^+ $ as the $2P(0^+)$ state. }\label{D3000plusresults}
	\vspace{0.3em}\centering
	\begin{tabular}{llccllc}
		\toprule[1.5pt]
		{\ Chanel }\qquad \quad \quad  & Final States \quad &\ \ \ 	Width \ \ \  & {\quad} & {Chanel }\qquad \quad \quad  & Final States \quad 	&\ \ \ 	Width \ \   \\
		\midrule[1pt]
		\multirow{2}{*}{ $ D(^1S_0) \pi $ }  & $ D^+ \pi^0 $ & 6.5 &  & \multirow{2}{*}{ $ D(2 ^1S_0)\pi$ } & $ D^0 \pi^+ $ &  3.8 \\
		&	$ D^0 \pi^+ $	&	13.5		&	& { } & $ D^0 \pi^+ $ & \ {7.7} \vspace{0.3em}\\
		{$  D \eta $}  & $ D^0 \eta^0 $    & 	0.56	&  	& {$  D \eta ' $}& $ D^0 \eta '^0$   & \  5.7 \vspace{0.3em} \\
		\multirow{2}{*}{$  D(2420)\pi$	}& $ D_1(2420)^+ \pi^0 $ & 18.3	&  & \multirow{2}{*}{$  D(2430) \pi$}& $ D_1(2430)^+ \pi^0 $ & 2.1 \\
		&  $ D_1(2420)^0 \pi^+ $ &  37.4	& & & $D_1(2430)^0 \pi^+$ & 4.3 \\
		{$  D(2420) \eta$}& $ D_1(2420)^+ \eta^0 $ & 0.77 &  & {$  D(2430) \eta$}& $ D_1(2430)^+ \eta ^0 $ & 0.11 \\	
		\multirow{2}{*}{$  D^{*} \rho$}& $ D^{*}(2010)^+ \rho ^0 $& 6.1 &   & {$  D^{*} \omega$}& $ D^{*}(2010)^+ \omega ^0 $ & 6.5 \\
		& $D^{*}(2007)^0 \rho ^-$ & 12.9 &   & {$  D_{s}(2460) K$}& $ D_{s1}(2460)^+ K^0 $ & 1.2 \\
		{$  D_s K$}	& $ D_s^+ K^0 $ & 0.05 &  & {$  D_s^{*} K^*$}& $ D_s^{*+} K^{*}(892)^0 $ & 3.8 \\
		\midrule[1pt]
		\multicolumn{3}{c}{Total} &\multicolumn{3}{c}{131.3} \\
		\bottomrule[1.5pt]
	\end{tabular}
\end{table}

Considering many theoretical prediction of the mass are lower than 3000 MeV \cite{Godfrey1985,Pierro2001,Ebert2009,Sun2013,Godfrey2016} and the properties of these states could be revised after more experimental data collected, we also calculate the total width changing with the mass from 2900 to 3020 MeV, which is shown in Fig.~\ref{width_mass_3000}. The total width of the neutral one ranges from 135.6 to 126.3 MeV, while the charged one's result changes from 136.8 to 127.7 MeV. The full width of the $2P$ state becomes narrower along with the phase space increasing, which is opposite to that of the $1P$ case. The reason is that the recoil momentum becomes more considerable when phase space is larger for the $2P$ state. This results in greater contribution from the part after the nodes, so the decay width gets smaller.
We also notice that there is a rise at the tail of the curve. It is because some new channels open when the mass of $D_J^*(3000)$ increase to 3000 MeV, such as $D_s^*K^*$, $D_1(2420)\eta$, $D_1(2430)\eta$. Thus, the total widths have the sudden rise.


\section{SUMMARY}

In this work, we study the two-body strong decay properties of two orbitally excited scalar $D$ mesons by the improved BS method. 
Our results of the $D_0^*(2400)$, as the $0^+(1P)$ states, are consistent with the present experimental data, which shows the suitability of our method. However, the sensitivity of decay width to its mass means more precise measurements are needed. 
For the $0^+(2P)$ assignment of $D_J^*(3000)$, the full decay width is about 131 MeV, which is a little higher but close to the present experimental data. Besides the $D\pi$ mode, we find $D\rho$ and $D_1(2420)\pi$ channels also contribute much to the full width, and they can be helpful in the further investigation. Considering the theoretical uncertainties from relativistic corrections of highly excited states and the preliminary experimental data at present, $D_J^*(3000)$ is still a strong candidate for the $2^3P_0$ state. We expect more experimental and theoretical efforts on this newly discovered resonance.


\section*{ACKNOWLEDGEMENTS}

This work was supported in part by the National Natural Science
Foundation of China (NSFC) under Grant No.~11575048, No.~11405037, No.~11505039, No.~11447601, No.~11535002 and No.~11675239.
We thank the HPC Studio at Physics Department of Harbin Institute of Technology for access to computing resources through INSPUR-HPC@PHY.HIT.

\appendix
\section*{APPENDIX  Bethe-Salpeter Wave Function}
\setcounter{equation}{0}
\renewcommand{\theequation}{\Alph{subsection}-\arabic{equation}}

BS method has been used extensively to describe the properties of heavy-light mesons \cite{WangGL2006,WangGL2009}. Here we only list some wave functions related to this work \cite{WangGL2004,lsc2017} 

\subsection{Wave function of $0^-$ state}

The general form of the wave function is 
\begin{equation}
\varphi_{0^-}(q_{\perp})=M\left[
\frac{\slashed{P}}{M}f_{b1}(q_\perp)+f_{b2}(q_\perp)+\frac{\slashed{q}_\perp}{M}f_{b3}(q_\perp)+\frac{\slashed{P}\slashed{q}_\perp}{M^2}f_{b4}(q_\perp)
\right] \gamma_5,
\end{equation}
where constraint conditions are 
\begin{equation}
\begin{split}
&f_{b3}=\frac{M(\omega_2-\omega_1)}{m_1\omega_2+m_2\omega_1} f_{b2}, \\
&f_{b4}=-\frac{M(\omega_1+\omega_2}{m_1\omega_2+m_2\omega_1}f_{b1}.
\end{split}
\end{equation}

The positive part is expressed as 
\begin{equation}
\varphi_{0^-}^{++}(q_{\perp}) = \left[ B_1(q_{\perp})+\frac{\slashed{P}}{M}B_2(q_{\perp})+\frac{\slashed{P} }{M}B_3(q_{\perp})+\frac{\slashed{P} \slashed{q}_{\perp}}{M^2}B_4(q_{\perp})
\right]\gamma _5.
\end{equation}

where
\begin{equation}
\begin{split}
&B_1\; =\frac{M}{2}\left(
\frac{\omega _1+\omega _2}{m_1+m_2}f_{b1}+f_{b2}
\right), \\
&B_2\; =\frac{M}{2}\left(
f_{b1}+\frac{m_1+m_2}{\omega _1 +\omega _2}f_{b2}
\right), \\
&B_3= -\frac{M(\omega _1 -\omega _2)}{m_1 \omega _2 +m_2 \omega _1}B_1, \\
&B_4 =- \frac{(m_1+m_2)M}{m_1 \omega _2+m_2 \omega _1}B_1.
\end{split}
\end{equation}

\subsection{Wave function of $1^+$ state}

We separate pseudo-vector $1^+$ state into $1^{++}$ and $1^{+-}$ to discuss.
The general form of $1^{++}(^3 P_1)$ state is 
\begin{equation}
\varphi_{1^{++}}(q_\perp)=\ui \varepsilon_{\mu \nu \alpha \beta}\frac{P^{\nu}}{M}q^{\alpha}_\perp \epsilon ^\beta \gamma^\mu \left[
\gamma ^\mu f_{c1}+\frac{\slashed{P}}{M}\gamma _\mu f_{c2}+\frac{q_\perp}{M} \gamma ^{\mu} f_{c3}+\frac{\slashed{P}\gamma ^\mu \slashed{q}_\perp}{M^2}f_{c4}
\right],
\end{equation}
where constraint conditions are 
\begin{equation}
\begin{split}
&f_{c3}=-\frac{M(\omega_1-\omega_2)}{m_1\omega_2+m_2\omega_1}f_{c1}, \\
&f_{c4}=\frac{M(\omega_1+\omega_2)}{m_1\omega_2+m_2\omega_1}f_{c2}.
\end{split}
\end{equation}

The positve part of the wave function is 
\begin{equation}
\varphi^{++}_{1^{++}}(q_\perp)=\ui \varepsilon_{\mu \nu \alpha \beta}\frac{P^{\nu}}{M}q^{\alpha}_\perp \epsilon ^\beta \gamma^\mu \left[
C_1(q_\perp)+\frac{\slashed{P}}{M}C_2(q_\perp)+\frac{q_\perp}{M}C_3(q_\perp)+\frac{\slashed{P}\slashed{q}_\perp}{M^2}C_4(q_\perp)
\right].
\end{equation}

where the conefficients are 
\begin{equation}
\begin{split}
&C_1\,=\frac{1}{2}\left(
f_{c1}+\frac{\omega_1+\omega_2}{m_1+m_2}f_{c2}
\right), \\
&C_2\,=-\frac{1}{2}\left(
\frac{m_1+m_2}{\omega_1+\omega_2}f_{c1}+f_{c2}
\right), \\
&C_3=\frac{M(\omega_1-\omega_2)}{m_1\omega_2+m_2\omega_1} C_1, \\
&C_4=-\frac{M(m_1+m_2)}{m_1\omega_2+m_2\omega_1} C_1.
\end{split}
\end{equation}

And the general form of $1^{+-}(^1 P_1)$ state is 
\begin{equation}
\varphi_{1^{+-}}(q_\perp)=q_\perp \cdot \epsilon \left[
f_{d1}(q_\perp)+\frac{\slashed{P}}{M}f_{d2}(q_\perp)+\frac{\slashed{q}_\perp}{M}f_{d3}(q_\perp)+\frac{\slashed{P}\slashed{q}_\perp}{M^2}f_{d4}(q_\perp)
\right] \gamma_5,
\end{equation}
Similar constraint condition is 
\begin{equation}
\begin{split}
f_{d3}=-\frac{M(\omega_1-\omega_2)}{m_1\omega_2+m_2\omega_1} f_{d1}, \\
f_{d4}=-\frac{M(\omega_1+\omega_2)}{m_1\omega_2+m_2\omega_1} f_{d2}.
\end{split}
\end{equation}

The postive part of the wave function is 
\begin{equation}
\varphi_{1^{+-}}^{++}(q_\perp)=q_\perp \cdot \epsilon \left[
D_1(q_\perp)+\frac{\slashed{P}}{M}D_2(q_\perp)+\frac{\slashed{q}_\perp}{M}D_3(q_\perp)+\frac{\slashed{P}\slashed{q}_\perp}{M^2}D_4(q_\perp)
\right] \gamma_5.
\end{equation}

where
\begin{equation}
\begin{split}
&D_1=\frac{1}{2}\left(
f_{d1}+\frac{\omega_1+\omega_2}{m_1+m_2}f_{d2}
\right),  \\
&D_2=\frac{1}{2}\left(
\frac{m_1+m_2}{\omega_1+\omega_2}f_{d1}+f_{d2}
\right), \\
&D_3=-\frac{M(\omega_1-\omega_2)}{m_1  \omega_2+m_2 \omega_1}D_1,  \\
&D_4=-\frac{M(m_1+m_2)}{m_1\omega_2+m_2\omega_1} D_1.
\end{split}
\end{equation}

\subsection{Wave function of $1^-$ state}

The wave function of $1^-(^3S_1)$ state is 
\begin{equation}
\begin{split}
\varphi_{1^-}(q_{\perp})= (q_{\perp}\cdot \varepsilon)\left[
f_{e1}(q_{\perp})+\frac{\slashed{P}}{M}f_{e2}(q_{\perp})+\frac{\slashed{q}_{\perp}}{M}f_{e3}(q_{\perp})+\frac{\slashed{P}\slashed{q}_{\perp}}{M^2}f_{e4}(q_{\perp})
\right] \\
+M\slashed{\varepsilon}\left[
f_{e5}(q_{\perp})+\frac{\slashed{P}}{M}f_{e6}(q_{\perp})+\frac{\slashed{q}_{\perp}}{M}f_{e7}(q_{\perp})+\frac{\slashed{P}\slashed{q}_{\perp}}{M^2}f_{e8}(q_{\perp})
\right].
\end{split}
\end{equation}

Constraint conditions are 
\begin{equation}
\begin{split}
f_{e1}=&\ \frac{q^2_{\perp}f_{e3}(\omega _1+\omega _2)+2M^2f_{e5}\omega _2}{M(m_1\omega _2+m_2\omega _1)}, \\
f_{e2}=&\ \frac{q^2_{\perp}f_{e4}(\omega _1-\omega _2)+2M^2f_{e6}\omega _2}{M(m_1\omega _2+m_2\omega _1)}, \\
&f_{e7}=\frac{M(\omega_1 - \omega_2}{m_1 \omega_2 + m_2\omega_1}f_{e5}, \\
&f_{e8}=\frac{M(\omega_1+\omega_2)}{m_1\omega_2+m_2\omega_1}f_{e6}.
\end{split}
\end{equation}

The positive part of the wave function is 
\begin{equation}
\begin{split}
\varphi_{1^-}^{++}(q_{\perp})=(q_{\perp}\cdot\varepsilon) \left[
E_{1}(q_{\perp})+\frac{\slashed{P}}{M}E_{2}(q_{\perp})+\frac{\slashed{q}_{\perp}}{M}E_3(q_{\perp})+\frac{\slashed{P}\slashed{q}_{\perp}}{M^2}E_4(q_{\perp})
\right] \\
+M \slashed{\varepsilon}\left[
E_5(q_{\perp})+\frac{\slashed{P}}{M}E_6(q_{\perp})+\frac{\slashed{q}_{\perp}}{M}E_7(q_{\perp})+\frac{\slashed{P}\slashed{q}_{\perp}}{M^2}E_8(q_{\perp})
\right],
\end{split}
\end{equation}

where the coefficients are 
\begin{equation}
\begin{split}
&E_1=\frac{1}{2M(m_1\omega_2+m_2\omega_1)}\left[
(\omega_1+\omega_2)q^2_{\perp}f_{e3}+(m_1+m_2)q^2_{\perp}f_{e4}+2M^2\omega_2 f_{e5}-2M^2 m_2 f_{e6}
\right], \\
&E_2 =\frac{1}{2M(m_1\omega_2+m_2\omega_1}\left[
(m_1-m_2)q^2_{\perp}f_{e3}+(\omega_1-\omega_2)q^2_{\perp}f_{e4}-2M^2 m_2 f_{e5}+2 M^2 \omega_2 f_{e6}
\right], \\
&E_3 =\frac{1}{2}\left[
f_{e3}+\frac{m_1+m_2}{\omega_1+\omega_2}f_{e4}-\frac{2 M^2}{m_1 \omega_2 + m_2 \omega_1} f_{e6}
\right], \\
&E_4 = \frac{1}{2}\left[
\frac{\omega_1 + \omega_2}{m_1 + m_2}f_{e3} + f_{e4} - \frac{2 M^2}{m_1 \omega_2 + m_2 \omega_1}f_{e5}
\right], \\
&E_5 = \frac{1}{2}\left[
f_{e5}-\frac{\omega_1 + \omega_2}{m_1 + m_2}f_{e6}
\right],\quad \quad
E_6 = \frac{1}{2}\left[
-\frac{m_1+m_2}{\omega_1+\omega_2}f_{e5}+f_{e6}
\right], \\
&E_7=\frac{M}{2}\frac{\omega_1-\omega_2}{m_1\omega_2+m_2\omega_1}\left[
f_{e5}-\frac{\omega_1+\omega_2}{m_1+m_2}f_{e6}
\right], \\
&E_8=\frac{M}{2}\frac{m_1+m_2}{m_1\omega_2+m_2\omega_1}\left[
-f_{e5}+\frac{\omega_1+\omega_2}{m_1+m_2}f_{e6}
\right].
\end{split}
\end{equation}

\bibliography{revised_version2.bib} 



\end{document}